\documentstyle[12pt]{article}
\begin{document}
\def \beq{\begin{equation}}
\def \eeq{\end{equation}}
\def \es{E$_{\rm 6}$}
\def \g{{\rm GeV}}
\def \nE{$\nu_E$}
\def \nbE{$\bar \nu_E$}
\renewcommand{\thetable}{\Roman{table}}
\rightline{DOE/ER/40561-61-INT99}
\rightline{EFI-99-33}
\rightline{hep-ph/9907438}
\vspace{0.5in}
\centerline{\bf MIXING OF CHARGE $-1/3$ QUARKS AND CHARGED LEPTONS}
\centerline{\bf WITH EXOTIC FERMIONS IN E$_{\rm 6}$
\footnote{To be submitted to Physical Review Letters}}
\vspace{0.5in}
\centerline{\it Jonathan L. Rosner}
\centerline{\it Institute for Nuclear Theory}
\centerline{\it University of Washington, Seattle, WA 98195}
\bigskip
\centerline{and}
\bigskip
\centerline{\it Enrico Fermi Institute and Department of Physics}
\centerline{\it University of Chicago, Chicago, IL 60637
\footnote{Permanent address.}}
\bigskip


\centerline{\bf ABSTRACT}
\medskip
\begin{quote}
A way is suggested to understand why the average masses of $(d,s,b)$ quarks are
smaller than those of $(u,c,t)$ quarks.  In contrast to previously proposed
mechanisms relying on different Higgs boson vacuum expectation values or
different Yukawa couplings, the mass difference is explained as a consequence
of mixing of $(d,s,b)$ with exotic quarks implied by the electroweak-strong
unification group E$_{\rm 6}$. 
\end{quote}
\bigskip

\leftline{PACS Categories: 12.10.Dm, 12.10.Kt, 12.60.Cn, 14.80.-j}
\bigskip


The currently known fermions consist of quarks $(u,c,t)$ with charge $2/3$,
quarks $(d,s,b)$ with charge $-1/3$, leptons $(e,\mu,\tau)$ with charge $-1$,
and neutrinos $(\nu_e,\nu_\mu,\nu_\tau)$. Some proposals address certain broad
features of their masses.  Specifically: 

\begin{itemize}

\item[{(1)}] The evidence \cite{solar,atm} that neutrino masses are non-zero
but tiny with respect to those of other fermions may be evidence for large
Majorana masses of right-handed neutrinos, which overwhelm Dirac mass terms and
lead to extremely small Majorana masses for left-handed neutrinos
\cite{seesaw}. 

\item[{(2)}] Many unified theories of the electroweak and strong interactions
(see, e.g.,~\cite{BEGN}) imply a relation between the masses of charged leptons
and quarks of charge $-1/3$ at the unification scale.  Such a relation does
seem to be approximately satisfied for the members $\tau,b$ of the heaviest 
family.

\item[{(3)}] The larger (average) masses of the $(u,c,t)$ quarks with respect
to the $(d,s,b)$ quarks could be a consequence of different vacuum expectation
values in a two-Higgs-doublet model \cite{Higgs}, where two different doublets
are responsible for the masses of quarks of different charges.  [In such a
picture we would view the masses of the lightest quarks, which have the
inverted order $m(d) > m(u)$, as due, for example, to a radiative effect, and
not characteristic of the gross pattern.]

\end{itemize}

In the present paper we propose another potential source of difference between
masses of quarks of different charges, which arises in a unified electroweak
theory based on the gauge group \es~\cite{E6refs,cmts}.  The fundamental
(27-dimensional) representation of this group contains additional quarks of
charge $-1/3$ and additional charged and neutral leptons, but no additional
quarks of charge $2/3$.  We have identified a simple mixing mechanism which can
depress the average mass of $(d,s,b)$ quarks (and charged leptons) with respect
to that of $(u,c,t)$ quarks without the need for different Higgs vacuum
expectation values.  This mixing can occur in such a way as to have minimal
effect on the weak charged-current and neutral-current couplings of quarks and
leptons, but offers the possibility of observable deviations from standard
couplings if the new states participating in the mixing are not too heavy. 
This mechanism was first observed in Ref.~\cite{ChM}. Similar mixing with
isosinglet quarks was discussed in Ref.~\cite{dAKQ}, but with a different
emphasis (including a mechanism for understanding $m_d > m_u$). A related
(``seesaw'') effect was used to describe the top quark mass in a particular
theory of electroweak symmetry breaking \cite{CDGH}. 

We first recall some basic features of \es~and mass matrices, and then describe
a scenario in which $(d,s,b)$ masses (and those of charged leptons) can be
depressed by mixing with their exotic \es~counterparts. Some consequences of
the mixing hypothesis are then noted. 

The fundamental 27-dimensional representation of \es~contains representations
of dimension 16, 10, and 1 of SO(10).  We assume there exist three 27-plets,
corresponding to the three quark-lepton families. We may regard ordinary matter
(including right-handed neutrinos) of a single quark-lepton family as residing
in an SO(10) 16-plet, with SU(5) content $5^* + 10 + 1$. The additional
(``exotic'') states in the 10-plet and singlet of SO(10) are summarized in
Table I for one family. Here $I_L$ and $I_{3L}$ refer to left-handed isospin
and its third component.

\begin{table}
\caption{Exotic fermions in a 27-plet of \es.}
\begin{center}
\begin{tabular}{c c c c c c} \hline \hline
SO(10) & SU(5) &  State  &  $Q$   & $I_L$ & $I_{3L}$ \\ \hline
10     & 5     &  $h^c$  & $1/3$  &   0   &   0    \\
       &       &  $E^-$  & $-1$   &  1/2  & $-1/2$ \\
       &       &   \nE   &   0    &  1/2  &   1/2  \\
       & $5^*$ &   $h$   & $-1/3$ &   0   &   0    \\
       &       &  $E^+$  &   1    &  1/2  &   1/2  \\
       &       &  \nbE   &   0    &  1/2  & $-1/2$ \\ \hline
1      &   1   &  $n_e$  &   0    &   0   &   0    \\ \hline \hline
\end{tabular}
\end{center}
\end{table}

All the new states are vector-like.  They consist of an isosinglet quark $h^c$
of charge $1/3$, a lepton isodoublet $(E^-,\nu_E)$, the corresponding
antiparticles, and a Majorana neutrino $n_e$. 

For simplicity we consider only mixings within a single family, which we shall
denote $(u,d,e,\nu_e)$. We shall discuss only mass matrices of charged
fermions.  The neutral lepton sector is of potential interest since it contains
possibilities for ``sterile'' neutrinos not excluded by the usual cosmological
and accelerator-based experimental considerations \cite{CM}. 

The simplest mass is that of the $u$ quark, which cannot mix with any other.
We can describe its contribution to the Lagrangian (we omit Hermitian
conjugates for brevity) in terms of a $2 \times 2$ matrix 
\beq
{\cal M}^u = \left[ \begin{array}{c c} 0 & m_u \\ m_u & 0 \end{array} \right]
\eeq
sandwiched between Weyl spinors $(u^c,u)$ and $(u^c,u)^T$.  The zeroes reflect
charge and baryon number conservation.  To diagonalize ${\cal M}^u$ it is most
convenient to square it and note that the corresponding eigenvalues $m_u^2$
come in pairs.  The simplest Higgs representation giving rise to $m_u$ belongs
to the $[27^*,10,5^*]$ of [\es, SO(10), SU(5)]. 

The corresponding mass matrix for quarks of charge $-1/3$ takes account of the
possible mixing between non-exotic $d$ and exotic $h$ quarks.  Its most general
form in a basis $(d^c,d,h^c,h)$ can be written \cite{cmts}
\beq \label{eqn:md}
{\cal M}^d = \left[ \begin{array}{c c c c}
  0  & m_2 &  0  & M_1 \\
 m_2 &  0  & m_3 &  0  \\
  0  & m_3 &  0  & M_2 \\
 M_1 &  0  & M_2 &  0  \\
 \end{array} \right]~~~.
\eeq
Here small letters refer to $\Delta I_L = 1/2$ masses, which are expected to be
of electroweak scale or less, while large letters refer to $\Delta I_L = 0$
masses, which can be of any magnitude (including the unification scale).  We
shall assume $m_i \ll M_i$.  If the masses in Eq.~(\ref{eqn:md}) arise through
vacuum expectation values of a Higgs $27^*$-plet (the simplest possibility),
their transformation properties are summarized in Table II.

\begin{table}
\caption{Simplest transformation properties of terms in ${\cal M}^d$.}
\begin{center}
\begin{tabular}{c c c} \hline \hline
Term  & SO(10) & SU(5) \\ \hline
$m_2$ &   10   &   5   \\
$m_3$ & $16^*$ &   5   \\
$M_1$ & $16^*$ &   1   \\
$M_2$ &    1   &   1   \\ \hline \hline
\end{tabular}
\end{center}
\end{table}

Eq.~(\ref{eqn:md}) is diagonalized, as before, by squaring it.  $({\cal
M}^d)^2$ decomposes into two separate $2 \times 2$ matrices, referring to the
bases $(d^c, h^c)$ and $(d,h)$.  For each of these, the eigenvalues
$\lambda_1$ and $\lambda_2$ satisfy
$$
\lambda_1 + \lambda_2 = m_2^2 + m_3^2 + M_1^2 + M_2^2~~,~~~
$$
\beq
\lambda_1 \lambda_2 = (M_1 m_3 - M_2 m_2)^2~~~.
\eeq

Suppose, to begin with, that $h$ and $h^c$ pair up to form a Dirac particle
with large mass $M_2 \gg (M_1,m_2,m_3)$.  Then the two eigenvalues are
$\lambda_1 \simeq m_2^2$ and $\lambda_2 \simeq M_2^2$, corresponding to light
and heavy Dirac particles $d$ and $h$, respectively. If we label basis states
with zeroes as subscripts, and physical states without subscripts, this
solution corresponds to $d = d_0$, $d^c = d^c_0$, $h = h_0$, $h^c = h^c_0$. 

For the more general case where $M_1$ is not negligible in comparison with
$M_2$, we can write
$$
M_1 = M \cos \theta~~,~~~M_2 = M \sin \theta~~,~~~
$$
\beq
m_3 = m \cos \phi~~,~~~m_2 = m \sin \phi~~~.
\eeq
Then for $m \ll M$, we have
\beq
\lambda_1 \simeq m^2 \cos^2(\theta + \phi)~~,~~~
\lambda_2 \simeq m^2 \sin^2(\theta + \phi) + M^2~~~.
\eeq
This is our central result.  It is possible to choose $\theta + \phi$
in such a way that the $d$ quark mass is arbitrarily small in comparison
with $m^2$, whose scale is a typical electroweak scale (as in the case of
$m^u$).  The opposite situation, in which $u$-type quarks are lighter than
$d$-type quarks, is unnatural in the present scheme.

The physical (left-handed) $(d^c,h^c)$ states are eigenstates of the matrix
\beq
{\cal M}^2_{d^c,h^c} = \left[ \begin{array}{c c}
m^2 \sin^2 \phi + M^2 \cos^2 \theta &
    m^2 \cos \phi \sin \phi + M^2 \cos \theta \sin \theta \\
m^2 \cos \phi \sin \phi + M^2 \cos \theta \sin \theta &
    m^2 \cos^2 \phi + M^2 \sin^2 \theta \\
\end{array} \right]~~~.
\eeq
For $M \gg m$ the approximate eigenstates are
\beq
d^c \simeq \sin \theta d^c_0 - \cos \theta h^c_0~~,~~~
h^c \simeq \cos \theta d_c^0 + \sin \theta h^c_0~~~.
\eeq
In the limit $\theta = \pi/2$ in which $M_2 \gg M_1$, leading to a large Dirac
mass for the exotic quark $h$, one thus has $d^c = d^c_0,~h^c = h^c_0$. 

The physical (left-handed) $(d,h)$ states are eigenstates of the matrix
\beq
{\cal M}^2_{d,h} = \left[ \begin{array}{c c}
m^2 & mM \sin(\theta + \phi) \\
mM \sin(\theta + \phi) & M^2 \\
\end{array} \right]~~~,
\eeq
specifically
\beq
d \simeq d_0 - (m/M) \sin (\theta + \phi) h_0~~,~~~
h \simeq (m/M) \sin (\theta + \phi) d_0 + h_0~~~.
\eeq
Thus, for $m \ll M$, there is little mixing between the isosinglet and
isodoublet quarks, and hence little potential for violation of unitarity of the
Cabibbo-Kobayashi-Maskawa (CKM) matrix.  Some consequences of this mixing have
been explored, for example, in Refs.~\cite{cmts,ChM,dAKQ,mix}. The mixing
parameter $\zeta \equiv (m/M) \sin(\theta+\phi)$ and the suppression of
$d$-type masses are both maximal for $\theta + \phi = \pm \pi/2$. 

Although the present mechanism for lowering the masses of down-type quarks
does not require $h$ quarks to be accessible at present energies, it is
interesting to speculate about this possibility. One effect of mixing between
an ordinary $d$-type quark and its exotic $h$-type counterpart is the
modification of couplings of the $b$ quark. The forward-backward asymmetry
$A_{FB}^b$ in $e^+ e^- \to Z \to b \bar b$, and the asymmetry parameter $A_b$
describing the couplings of the $b$ to the $Z$, are slightly different from the
values expected in a standard electroweak fit, where 
\beq
g_{bL} = - \frac{1}{2} + \frac{1}{3}\sin^2 \theta_W~~,~~~
g_{bR} = \frac{1}{3} \sin^2 \theta_W~~~,
\eeq
and $A_b = (g_{bL}^2 - g_{bR}^2)/ (g_{bL}^2 + g_{bR}^2)$ is predicted to be
0.935 for $\sin^2 \theta_W = 0.2316$.  To account for the experimental value
\cite{Chan,Ab} of $A_b = 0.891 \pm 0.017$ while keeping $g_{bL}^2 + g_{bR}^2$
fixed (since the total rate for $Z \to b \bar b$ is now in accord with standard
model predictions) one must modify both $g_{bL}$ and $g_{bR}$ in such a way
that $g_{bL} \delta g_{bL} \simeq - g_{bR} \delta g_{bR}$ \cite{Chan}. 

The present scheme does not fill the bill, since it affects only left-handed
couplings, mixing an isodoublet $d$ with an isosinglet $h$. We find $\delta
g_{bL} = \zeta^2/2$, $\delta g_{bR} = 0$.  Here we have assumed an unmixed $Z$.
Severe constraints apply to the mixing of the $Z$ with a higher-mass $Z'$
\cite{EL}.

The production of $h \bar h$ pairs in hadronic collisions should be governed by
standard perturbative QCD, which gives a reasonable account of top quark pair
production \cite{topprod}.  For the data sample of approximately 100 pb$^{-1}$
obtained in $p \bar p$ collisions at a center-of-mass energy of 1.8 TeV in Run
I at the Fermilab Tevatron, it should be possible to observe or exclude values
of $m(h)$ well in excess of $m(t)$ \cite{Andre}.  It may also be possible to
produce or exclude $h$ quarks singly through the neutral flavor-changing
interaction at LEP II via the reaction $e^+ e^- \to Z^* \to h + (\bar d, \bar
s, \bar b)$. Both charged-current decays $h \to W + (u,c,t)$ and
neutral-current decays $h \to Z + (d,s,b)$ should be characterized by multiple
leptons and missing energy in an appreciable fraction of events. 

The mixing proposed here applies in an almost identical manner to the charged
leptons under the replacements $d^c \to e^-$, $d \to e^+$, $h^c \to E^-$, $h
\to E^+$.  The charged leptons' masses, just like those of the $d$-type quarks,
thus may be depressed relative to their unmixed values. One could expect small
modifications of {\it right-handed} lepton couplings since one is then mixing
an isosinglet $e$ with an isodoublet $E$. 

To conclude, we have presented a mechanism which accounts for the depression
in the average masses of down-type quarks and charged leptons relative to that
of up-type quarks, without the need for differences in Higgs vacuum expectation
values or in values of the largest Yukawa coupling for each type of fermion.
This mechanism relies on mixings between ordinary fermions and their exotic
counterparts in \es~multiplets.  It may be of use in building more realistic
models of quark and lepton masses.  Although the exotic \es~fermions need not
be accessible to present experimental searches in order for this mechanism to
be effective, they could well be observable in forthcoming searches at the
Fermilab Tevatron, the LEP II $e^+ e^-$ collider, or the Large Hadron Collider
under construction at CERN.

I am indebted to T. Andr\'e, F. del Aguila, B. Kayser, R. N. Mohapatra, S. T.
Petcov, and L. Wolfenstein for useful discussions. I wish to thank the
Institute for Nuclear Theory at the University of Washington for hospitality
during this work, which was supported in part by the United States Department
of Energy under Grant No. DE FG02 90ER40560. 
\bigskip

\def \ajp#1#2#3{Am. J. Phys. {\bf#1}, #2 (#3)}
\def \apny#1#2#3{Ann. Phys. (N.Y.) {\bf#1}, #2 (#3)}
\def \app#1#2#3{Acta Phys. Polonica {\bf#1}, #2 (#3)}
\def \arnps#1#2#3{Ann. Rev. Nucl. Part. Sci. {\bf#1}, #2 (#3)}
\def \cmts#1#2#3{Comments on Nucl. Part. Phys. {\bf#1}, #2 (#3)}
\def \cn{Collaboration}
\def \cp89{{\it CP Violation,} edited by C. Jarlskog (World Scientific,
Singapore, 1989)}
\def \dpfa{{\it The Albuquerque Meeting: DPF 94} (Division of Particles and
Fields Meeting, American Physical Society, Albuquerque, NM, Aug.~2--6, 1994),
ed. by S. Seidel (World Scientific, River Edge, NJ, 1995)}
\def \dpff{{\it The Fermilab Meeting: DPF 92} (Division of Particles and Fields
Meeting, American Physical Society, Batavia, IL., Nov.~11--14, 1992), ed. by
C. H. Albright \ite~(World Scientific, Singapore, 1993)}
\def \efi{Enrico Fermi Institute Report No. EFI}
\def \epj#1#2#3{Eur.~Phys.~J.~{\bf #1}, #2 (#3)}
\def \epl#1#2#3{Europhys.~Lett.~{\bf #1}, #2 (#3)}
\def \f79{{\it Proceedings of the 1979 International Symposium on Lepton and
Photon Interactions at High Energies,} Fermilab, August 23-29, 1979, ed. by
T. B. W. Kirk and H. D. I. Abarbanel (Fermi National Accelerator Laboratory,
Batavia, IL, 1979}
\def \hb87{{\it Proceeding of the 1987 International Symposium on Lepton and
Photon Interactions at High Energies,} Hamburg, 1987, ed. by W. Bartel
and R. R\"uckl (Nucl. Phys. B, Proc. Suppl., vol. 3) (North-Holland,
Amsterdam, 1988)}
\def \ib{{\it ibid.}~}
\def \ibj#1#2#3{~{\bf#1}, #2 (#3)}
\def \ichep72{{\it Proceedings of the XVI International Conference on High
Energy Physics}, Chicago and Batavia, Illinois, Sept. 6 -- 13, 1972,
edited by J. D. Jackson, A. Roberts, and R. Donaldson (Fermilab, Batavia,
IL, 1972)}
\def \ijmpa#1#2#3{Int. J. Mod. Phys. A {\bf#1}, #2 (#3)}
\def \ite{{\it et al.}}
\def \jpb#1#2#3{J.~Phys.~B~{\bf#1}, #2 (#3)}
\def \lkl87{{\it Selected Topics in Electroweak Interactions} (Proceedings of
the Second Lake Louise Institute on New Frontiers in Particle Physics, 15 --
21 February, 1987), edited by J. M. Cameron \ite~(World Scientific, Singapore,
1987)}
\def \ky85{{\it Proceedings of the International Symposium on Lepton and
Photon Interactions at High Energy,} Kyoto, Aug.~19-24, 1985, edited by M.
Konuma and K. Takahashi (Kyoto Univ., Kyoto, 1985)}
\def \mpla#1#2#3{Mod. Phys. Lett. A {\bf#1}, #2 (#3)}
\def \nc#1#2#3{Nuovo Cim. {\bf#1}, #2 (#3)}
\def \ncl#1#2#3{Lettere al Nuovo Cim. {\bf#1}, #2 (#3)}
\def \np#1#2#3{Nucl. Phys. {\bf#1}, #2 (#3)}
\def \pisma#1#2#3#4{Pis'ma Zh. Eksp. Teor. Fiz. {\bf#1}, #2 (#3) [JETP Lett.
{\bf#1}, #4 (#3)]}
\def \pl#1#2#3{Phys. Lett. {\bf#1}, #2 (#3)}
\def \pla#1#2#3{Phys. Lett. A {\bf#1}, #2 (#3)}
\def \plb#1#2#3{Phys. Lett. B {\bf#1}, #2 (#3)}
\def \pr#1#2#3{Phys. Rev. {\bf#1}, #2 (#3)}
\def \prc#1#2#3{Phys. Rev. C {\bf#1}, #2 (#3)}
\def \prd#1#2#3{Phys. Rev. D {\bf#1}, #2 (#3)}
\def \prl#1#2#3{Phys. Rev. Lett. {\bf#1}, #2 (#3)}
\def \prp#1#2#3{Phys. Rep. {\bf#1}, #2 (#3)}
\def \ptp#1#2#3{Prog. Theor. Phys. {\bf#1}, #2 (#3)}
\def \rmp#1#2#3{Rev. Mod. Phys. {\bf#1}, #2 (#3)}
\def \rp#1{~~~~~\ldots\ldots{\rm rp~}{#1}~~~~~}
\def \si90{25th International Conference on High Energy Physics, Singapore,
Aug. 2-8, 1990}
\def \slc87{{\it Proceedings of the Salt Lake City Meeting} (Division of
Particles and Fields, American Physical Society, Salt Lake City, Utah, 1987),
ed. by C. DeTar and J. S. Ball (World Scientific, Singapore, 1987)}
\def \slac89{{\it Proceedings of the XIVth International Symposium on
Lepton and Photon Interactions,} Stanford, California, 1989, edited by M.
Riordan (World Scientific, Singapore, 1990)}
\def \smass82{{\it Proceedings of the 1982 DPF Summer Study on Elementary
Particle Physics and Future Facilities}, Snowmass, Colorado, edited by R.
Donaldson, R. Gustafson, and F. Paige (World Scientific, Singapore, 1982)}
\def \smass90{{\it Research Directions for the Decade} (Proceedings of the
1990 Summer Study on High Energy Physics, June 25--July 13, Snowmass, Colorado),
edited by E. L. Berger (World Scientific, Singapore, 1992)}
\def \tasi90{{\it Testing the Standard Model} (Proceedings of the 1990
Theoretical Advanced Study Institute in Elementary Particle Physics, Boulder,
Colorado, 3--27 June, 1990), edited by M. Cveti\v{c} and P. Langacker
(World Scientific, Singapore, 1991)}
\def \yaf#1#2#3#4{Yad. Fiz. {\bf#1}, #2 (#3) [Sov. J. Nucl. Phys. {\bf #1},
#4 (#3)]}
\def \zhetf#1#2#3#4#5#6{Zh. Eksp. Teor. Fiz. {\bf #1}, #2 (#3) [Sov. Phys. -
JETP {\bf #4}, #5 (#6)]}
\def \zpc#1#2#3{Zeit. Phys. C {\bf#1}, #2 (#3)}
\def \zpd#1#2#3{Zeit. Phys. D {\bf#1}, #2 (#3)}

\end{document}